\documentclass[12pt,a4paper]{article}
\usepackage{amssymb}
\usepackage{amsfonts}
\usepackage{amsmath}

\input{tcilatex}

\begin{document}

\title{Complete set of unitary irreps of Discrete Heisenberg Group $%
HW_{2^{s}}$.}
\author{E. Floratos$^{1,2}$ and I. Tsohantjis$^{1}$ \thanks{%
For correspondence: I. Tsohantjis, itsochantzis@phys.uoa.gr} \\
$^1$National and Kapodistrian University of Athens, \\
Faculty of Physics, Department of Nuclear and Particle Physics,\\
University Campus, Ilisia, 15771 Athens, Greece\\
$^2$Academy of Athens, Panepistimiou 28, 106 79, Athens, Greece}
\maketitle

\begin{abstract}
Following the method of induced group representations of Wigner-Mackay, the
explicit construction of all the unitary irreducible representations of the
discrete finite Heisenberg-Weyl group $HW_{2^{s}}$ over the discrete phase
space lattice $\mathbb{Z}_{2^{s}}\otimes $ $\mathbb{Z}_{2^{s}}$ is
presented. We explicitly determine their characters and their fusion rules.
We discuss possible physical applications for finite quantum mechanics and
quantum computation.
\end{abstract}



\vspace{10pt} 

\section{Introduction}

Quantum computation and quantum information are fast advancing areas of
theoretical and experimental studies of a variety of physical systems,
ranging from integrable lattice models, many body systems, anyonic systems
to black holes.

Technological advances in the control and study of few atoms quantum systems
require the isolation of these systems from the environment. Restricting and
isolating the study to transitions between a fixed finite number of energy
levels reduces the quantum mechanical problem to finite dimensional Hilbert
spaces.

The usual theoretical approach is the description of the quantum systems
with a finite number of spins and the corresponding models are various forms
of one or two dimensional Heisenberg spin chains. On the other hand in this
picture the role of the discrete Heisenberg-Weyl group which is the
mathematical expression of wave-particle duality is not obvious. The
Heisenberg-Weyl group plays an important role also in quantum information on
phase space qubits \cite{woot,woot2} as well as in quantum error correcting
codes \cite{shor,howe}.

In another area that of the fundamental questions about the nature of
gravity and spacetime there exist great difficulties in the unification of
gravity with quantum mechanics. It is becoming obvious that due to the
assignment of finite entropy to every local causal region connecting two
different observers, the corresponding Hilbert space of states must be
finite dimensional (Refs. \cite{Banks, caroll, carroll2} and references
therein).

In the case where the dimension of the Hilbert space is $2^{s}$ which
relevant for quantum computing and more generally for Heisenberg spin
chains, typically one starts with a quantum system of $s$ qubits.
Information processing on such systems is carried out through quantum
circuits which represent the action of unitary operators, $2^{s}$ $\times
2^{s}$ matrices acting on $H_{2^{s}}$. These operators are themselves
constructed as a tensor product of a finite number of one, two or three
qubit unitary operators, the universal quantum gates, and describe the
quantum mechanical evolution of the $s$-qubit system from an initial to a
final state.

In this standard picture of quantum information processing the fundamental
character of quantum mechanics namely the wave-particle duality is only
implicit and appears to play only a minor role if not at all. This situation
is completely different to that of the continuous quantum mechanical systems
where wave-particle duality and Heisenberg uncertainty principle play a
central role in the description of the evolution of a quantum system. To
this end one should focus in the inherently discrete character of
information processing of $s$-qubits quantum systems where the wave-particle
duality is expressed through the fundamental Heisenberg Weyl uncertainty
relation at the group level: $QP=\omega PQ$ , where $\omega =e^{2\pi
i/2^{s}} $ , $Q$ is the position operator $Q=diag[1,\omega ,\omega
^{2},...,\omega ^{2^{s}-1}]$ for a fictitious particle hoping(anticlockwise)
on the vertices of a canonical $s$-agon situated on the unit circle, and $P$
is the discrete momentum operator performing this (anticlockwise) motion by
an $2\pi /2^{s}$-angle shift on the circle.

In this picture the wave-particle duality is transparent. The wave functions
which are the eigenstates of $P$ are the discretization of the continuous
planewaves. These eigenstates are the columns of a diagonalization matrix $F 
$ of $P$, with matrix elements $F_{kj}=1/\sqrt{2^{s}}\omega ^{kj}$\
,satisfying the relations $QF=FP$, $F^{4}=I$. $F$ is thus a Finite Fourier
transform between discrete position and momentum of the fictitious
particle's phase space $\mathbb{Z}_{2^{s}}\times \mathbb{Z}_{2^{s}}$.

The general framework of Finite Quantum Mechanics (FQM), is quantum
mechanics with a finite dimensional Hilbert space. The reason for
introducing this framework has been exemplified and discussed in detail by 
\cite{weyl,swinger,balian,vourdas,jagan}. We should stress though that the
previous references concern with the fundamentals of quantum mechanics for
physical systems and there is no discussion about applications of this
framework in quantum information.

In the present work we shall connect the area of quantum information with
that of FQM. To this end, in this article, we first determine all the
inequivalent irreducible unitary representations of the discrete
Heisenberg-Weyl group $HW_{2^{s}}$ as well as their properties. This is the
necessary first step to define FQM on the discrete phase space $%
\mathbb{Z}
_{2^{s}}\times 
\mathbb{Z}
_{2^{s}}$.The unitary evolution operators of FQM will be obtained from the
unitary metaplectic representations of $Sp(2,%
\mathbb{Z}
_{2^{s}})$\ which is the automorphisms group of $HW_{2^{s}}$ \cite{flo8}.
The plan of the paper is as follows:

In section 2 the structure of $HW_{2^{s}}$ together with the determination
of its conjugacy classes is presented. In this paper we adopted a somewhat
more abstract generalized definition of this group in terms of $%
\mathbb{Z}
_{2^{s}}$ central extention of $%
\mathbb{Z}
_{2^{s}}\times 
\mathbb{Z}
_{2^{s}}$ intead of being defined for example as 3x3 upper triangular
matrices with 1 in the diagonal and the rest entries taking values in the $%
\mathbb{Z}
/2^{s}%
\mathbb{Z}
$. In any case it is the Heisenberg-Weyl group $HW_{2^{s}}$ over the ring $%
\mathbb{Z}
/2^{s}%
\mathbb{Z}
$ we examine as in this way we discover new representations relevant for
quantum mechanical computation and information.

In section 3 we derive up to equivalence all the inequivalent unitary
irreducible representations of the discrete Heisenberg group $HW_{2^{s}}$,
using Wigner's and MacKey's little group method \ on induced representations
for the case of a semi-direct product group. Note though that since we use
the abstract generalised definition of the group through generators and
their relations, we do not restrict ourselves to only primitive roots of
unity $\omega $ but also for any other admissible even or odd power of $%
\omega $. This corresponds to quantum mechanics with various Planck
constants $h=2\pi k/2^{s}$, where $k=1,...,2^{s}-1$. In this way we obtain
not only the faithful representations but also non-faithful ones \cite%
{zak,grass}. It seems that the non faithful ones have not attracted much
attention until now in the literature. Representations of discrete
Heisenberg group have in one form or another, appeared in the literature in
different context and areas of applications \cite{roditi, misag,schulte,
gure, apleby}. It appears that Ref. \cite{schulte} is closest to our
presentation although the important problem of fusion rules is not
discussed. On the other hand Ref. \cite{gure} although being a
mathematically coherent presentation of representations of Heisenberg group
over the ring $%
\mathbb{Z}
/2^{d}%
\mathbb{Z}
$ it lucks explicit derivation of the tools needed for direct computation of
physical quantities. It is our analysis that aims to an explicit derivation
of all matrix irreps, character sets and their fusion rules to be
implemented on physical applications mentioned above. That is why the
Wigner's and MacKey's method has been used in this article, as this method
has been proved to be very transparent when one describes finite group
symmetries (and not only) in physical systems.

In section 4 we obtain the explicit matrix form of constructed
representations which is then used to determine Finite Fourier Transforms
and their properties corresponding both for the case of the faithful and
non-faithful representations.

In section 5, the characters of the above representations are constructed as
well as the decomposition of direct product of any two inequivalent
representations is carried out in detail, thus establishing the set of
fussion rules for $HW_{2^{s}}$. We present examples of fussion rules for $%
HW_{2}$ and $HW_{4}$

In section 6 we conclude and we discuss open issues and possible
applications.

Finally in the appendix Section A, we give all the details for the
determination of the distinct orbits of the little group methods.

\section{Structure of the discrete and finite Heisenberg-Weyl group $HW_{N}$}

In this section we define the discrete and finite Heisenberg-Weyl group $%
HW_{N}$ and determine its conjugacy classes.

$HW_{N}$ is the set of elements $z^{m}x^{n}y^{l}$ where $m,n,l=0,1,...N-1$
and $x$, $\ y$, $z$ are subject to the relations 
\begin{eqnarray}
x^{N} &=&y^{N}=z^{N}=e,\;yx=zxy,\;  \label{def} \\
zx &=&xz,\;zy=yz\text{ \ \ \ \ \ }
\end{eqnarray}%
\ \ $e$ is the group identity element and the group has order $%
|HW_{N}|=N^{3} $. $HW_{N}$ has the semi-direct product group structure 
\begin{equation}
HW_{N}=\mathit{H}\circledS \mathit{B}  \label{struct}
\end{equation}%
where the abelian subgroups $\mathit{H}$ and $\mathit{B}$ are given by 
\begin{equation}
\mathit{H}=\{z^{m}x^{n}|m,n=0,1...N-1\}  \label{defH}
\end{equation}%
\begin{equation}
\mathit{B}=\{y^{l}|l=0,1...,N-1\}\approx 
\mathbb{Z}
_{N}.  \label{defB}
\end{equation}%
It can be shown that if the prime decomposition of $N$ is 
\begin{equation}
N=p_{1}^{r_{1}}p_{2}^{r_{2}}...p_{M}^{r_{M}},  \label{primedec}
\end{equation}%
where $p_{1},p_{2},...,p_{M}$ distinct primes and $r_{1},r_{2},...,r_{M}$
positive integers, then $HW_{N}$ admits the tensor product structure\cite%
{swinger} 
\begin{equation}
HW_{N}=\otimes _{i=1}^{M}HW_{p_{i}^{r_{i}}}.  \label{struct2}
\end{equation}%
Due to the structure (\ref{struct2}), it only suffices to find all the
unitary inequivalent irreps of $HW_{N}$ for $N=p^{s}$, $p$ prime and $s$
positive integer.

In this article we shall focus on the specific case where $N=2^{s}$. The
general case for $N=p^{s}$, $p$ prime will be dealt in future article 

We proceed to the determination of the conjugacy classes of $HW_{2^{s}}$.
Let $g=z^{\beta }x^{n}y^{l}$ , for all $n$, $l$, $\beta \in 
\mathbb{Z}
_{2^{s}}$, be an arbitrary element of the group. We shall discuss with
increasing complexity the conjugacy class of $g$
. The distinct elements of the conjugacy class $C_{g}$ are obtained by
conjugation $hgh^{-1}$ with all elements $h=z^{a}x^{\rho }y^{m}$ of $%
HW_{2^{s}}$, $a$, $\rho $, $m\in 
\mathbb{Z}
_{2^{s}}$, and will have the form $z^{mn-\rho l+\beta }x^{n}y^{l}$, for
particular values of $m$, $\rho $ to be found.

First we consider the case $n=l=0$. It is obvious that for each value of $%
\beta =0,1,...2^{s}-1$ the corresponding conjugacy class is $C_{z^{\beta
}}=\{z^{\beta }\}$ and so in total we have $2^{s}$ conjugacy classes of one
element each.

Secondly we consider the case $n\neq 0$, $l=0$, i.e. elements of the form $%
g=z^{\beta }x^{n}$. In this case the conjugacy class contains elements of
the form $z^{mn+\beta }x^{n}$ where $m=0,1,2,...,2^{s}-1$. However, for
fixed $n$, $\beta $, not all of these elements are different . To count the
distinct ones we consider the problem of finding the distinct points on the
discrete line of the $(j,m)-$plane, where $j=(mn+\beta )$ ${mod}2^{s}$, as $m
$ take the values $0,1,2,...,2^{s}-1$. To solve this problem we extract the
maximum power of $2$ from $n$ by expressing $n=2^{t}u_{t}$, where $%
t=0,1,2,...,s-1$ and $u_{t}=1,3,5..,2^{s-t}-1,$ and thus $%
j=(2^{t}u_{t}m+\beta ){mod}2^{s}$. Direct investigation shows that for fixed 
$\beta =0,1,2,...,2^{t}-1,$ the values of $j$, when $m=0,...,2^{s-t}-1$ are
given by $j=\beta ,2^{t}u_{t}+\beta ,...,(2^{s-t}-1)2^{t}u_{t}$ $+\beta $
and are respectivelly repeated for $m=2^{s-t}$,$2^{s-t}+1,$...,$2\times
2^{s-t}-1,$ and again for $m=2\times 2^{s-t}$, $2\times 2^{s-t}+1,$...,$%
2\times 2^{s-t}+2^{s-t}-1$ and so on, till $m=(2^{t}-1)\times 2^{s-t}$, $%
(2^{t}-1)\times 2^{s-t}+1,...,2^{s}-1)$. Moreover for the values of $m=0$,..,%
$2^{s-t}-1$, and $u_{t}\ $\ as above, it can be easily seen that the term $%
\gamma =2^{t}u_{t}m$ $({mod}2^{s})$ takes the values $\gamma
=0,2^{t},2\times 2^{t},...,(2^{s-t}-1)2^{t}$ and thus $j=$ $\gamma +\beta $.
However for $\beta =2^{t},2^{t}+1,...,2\times 2^{t}-1$ the values of $j$ are
respectively repeated as for $\beta =0,1,2,...,2^{t}-1$, and the same occurs
for $\beta =2\times 2^{t},2\times 2^{t}+1,...,3\times 2^{t}-1$, and so on,
until $\beta =(2^{s-t}-1)\times 2^{t},(2^{s-t}-1)\times 2^{t}+1,...,2^{s}-1$%
. Consequently the distinct conjugacy classes with distinct elements
obtained for $g=z^{\beta }x^{n}$ are of the form $C_{z^{\beta
}x^{n}}=\{z^{\gamma +\beta }x^{n},$ $\gamma =0,2^{t},2\times
2^{t},...,(2^{s-t}-1)2^{t}\},$ where the values of $\beta $ are restricted
as above.The number of conjugacy classes thus obtained is given by $%
\sum_{t=0}^{s-1}2^{t}(2^{s-t-1})$ each one containing $2^{s-t}-1$ elements.

Thirdly, in exactly the same way as above we obtain the conjugacy classes
for the cases $g=z^{\beta }y^{l}$, with $l=2^{t^{\prime }}u_{t^{\prime }}$, $%
t^{\prime }=0,1,2,...,s-1$ and $u_{t^{\prime }}=1,3,5..,2^{s-t^{\prime }}-1$%
, to be $C_{z^{\beta }y^{l}}=\{z^{\gamma ^{\prime }+\beta }y^{l},$ $\gamma
^{\prime }=0,2^{t^{\prime }},2\times 2^{t^{\prime }},...,(2^{s-t^{\prime
}}-1)2^{t^{\prime }}$ $\}$ and where the values of $\beta $ are again
restricted as above and the number of conjugacy classes is given by $%
\sum_{t^{\prime }=0}^{s-1}2^{t^{\prime }}(2^{s-t^{\prime }-1})$ of $%
2^{s-t^{\prime }}-1$ elements each.

Finally, when $g=z^{\beta }x^{n}y^{l}$ with $n=2^{t}u_{t}$, $l=2^{t^{\prime
}}u_{t^{\prime }},$ we follow a similar investigation line. In this case $%
j=(2^{t}u_{t}m-2^{t^{\prime }}u_{t^{\prime }}\rho +\beta ){mod}2^{s}$ and
for fixed $\beta $ the values of $j$ are those for which $%
2^{t}u_{t}m-2^{t^{\prime }}u_{t^{\prime }}\rho +\beta <2^{s}$. When $%
t^{\prime }\geq t$ \ we exress $j=[2^{t}u_{t}(m-2^{t^{\prime }-t}\frac{%
u_{t^{\prime }}}{u_{t}}\rho )+\beta ]{mod}2^{s}$ and the distinct values of $%
j$ are obtained when the term $(m-2^{t^{\prime }-t}\frac{u_{t^{\prime }}}{%
u_{t}}\rho ){mod}2^{s}$ is restricted to the values $0,1,...,(2^{s-t}-1)$
while the values of $\beta $ are restricted to $0,1,2,...,2^{t}-1$. In this
case the number of conjugacy classes is given by $\sum_{t,t^{\prime
}=0,t^{\prime }\geq t}^{s-1}2^{2s-t-t^{\prime }-2}2^{t}$. Equivalently, when 
$t>t^{\prime }$, we exress $j=[2^{t^{\prime }}u_{t^{\prime }}(2^{t-t^{\prime
}}\frac{u_{t}}{u_{t^{\prime }}}m-\rho )+\beta ]{mod}2^{s}$ and the distinct
values of $j$ are obtained when the term $(2^{t-t^{\prime }}\frac{u_{t}}{%
u_{t^{\prime }}}m-\rho ){mod}2^{s}$ is restricted to the values $%
0,1,...,(2^{s-t^{\prime }}-1)$ while now the values of $\beta $ are
restricted to $0,1,2,...,2^{t^{\prime }}-1$ and the number of conjugacy
classes is $\sum_{t,t^{\prime }=0,t>t^{\prime }}^{s-1}2^{2s-t-t^{\prime
}-2}2^{t}$. It is thus evident that direct analysis of the distinct values
of $(m-2^{t^{\prime }-t}\frac{u_{t^{\prime }}}{u_{t}}\rho )$ and $%
(2^{t-t^{\prime }}\frac{u_{t}}{u_{t^{\prime }}}m-\rho )$ shows that they can
be put in the form $0,2^{k},2\times 2^{k},3\times 2^{k},...,(2^{s-k}-1)2^{k}$%
, where $k=\min (t,t^{\prime })$, while the values of $\beta $ are
restricted to $0,1,2,...,2^{k}-1$. Consequently, the conjugacy classes
obtained in this case are given by $C_{z^{\beta }x^{n}y^{l}}=\{z^{\mu +\beta
}x^{n}y^{l}, \mu=0,2^{k},2\times 2^{k},...,(2^{s-k}-1)2^{k} \}$ and the
number of conjugacy classes thus obtained is $\sum_{t,t^{\prime
}=0}^{s-1}(2^{s-t-1}2^{s-t^{\prime }-1}2^{k})$ each one containing $%
2^{s-k}-1 $ elements.

Collecting all of the above cases, the conjugacy classes of $HW_{2^{s}}$ are
of the following form:

\begin{equation}
C_{z^{\beta }x^{n}y^{l}}=\{z^{\alpha +\beta }x^{n}y^{l},\alpha
=0,2^{k},2\times 2^{k},3\times 2^{k},...,(2^{s-k}-1)2^{k}\}  \label{con}
\end{equation}%
where $n$, $l\in \{0,1,...,2^{s}-1\}$, $n=2^{t}u_{t}$, $l=2^{t^{\prime
}}u_{t^{\prime }}^{\prime }$, $t,t^{\prime }\in \{0,1,2,...,s\}$, $%
u_{t}=1,3,5..,2^{s-t}-1$, $u_{t^{\prime }}^{\prime }=1,3,5..,2^{s-t^{\prime
}}-1$, $k=\min (t,t^{\prime })$, $u_{s}=0$ and $\beta $ must be restricted
to the range $\{0,1,...,2^{k}-1\}$ in order to avoid double counting. The
order of each $C_{z^{\beta }x^{n}y^{l}}$ is $|C_{z^{\beta
}x^{n}y^{l}}|=2^{s-k}$. Counting arguments show that the total number of
conjugacy classes $N_{C}$ is given by 
\begin{equation}
N_{C}=\sum_{t,t^{\prime }=0}^{s-1}(2^{2s-t-t^{\prime }-2}2^{\min
(t,t^{\prime })})+2^{s}(s+1)  \label{numcon}
\end{equation}

In what follows, we shall derive the complete set of unitary irreducible
matrix representations of $HW_{2^{s}}$ using the method of induced
representations as applied to finite groups and their number must be equal
to $N_{C}$ above.

\section{Construction of the complete set of unitary irreducible matrix
representations of $HW_{2^{s}}.$}

In this section, we will present Induced representations are provided by the
following theorem (c.f. Ref.~\cite{mack,corn}). Bellow we will merely follow
Ref.~\cite{corn}.

\textit{Theorem 1} Let $G$ be a finite group and $K$ a subgroup of $G$, of
respective orders $|G|$ and $|K|$. Let $\Delta $ be a $d$-dimensional
representation of $K$ . Let $g_{i}$, $i=1,2,...,\frac{|G|}{|K|}$ be the
coset representatives for the decomposition of $G$, in to right cosets wrt $%
K $. Then the set of $d\frac{|G|}{|K|}\times d\frac{|G|}{|K|}$ matrices $%
\Gamma (g)$ defined by 
\begin{eqnarray}
\Gamma (g)_{im,jn} &=&\Delta (g_{i}gg_{j}^{-1})_{mn}\text{ \ \ for all \ \ }%
g_{i}gg_{j}^{-1}\in K  \notag \\
&=&0\text{ \ \ \ \ \ \ \ \ \ \ \ \ \ \ \ \ \ \ otherwise}  \label{definduce}
\end{eqnarray}%
for all $g\in G$, \ where $i,j=1,2,...,\frac{|G|}{|K|}$ and $m,n=1,..,d$,
provide a $d\frac{|G|}{|K|}$-dimensional representation of $G$ induced from
the $d$-dimensional representation $\Delta $ of $K$. For the induced
representation $\Gamma $ of a group $G,$ obtained from a representation $%
\Delta $ of one of its subgroups $K,$ we formally denote as $\Gamma =\Delta
(K)\uparrow G$. Moreover the characters $\chi $\ of the representation $%
\Gamma $\ \ are given by 
\begin{equation}
\chi (g)=\sum_{i=1}^{\frac{|G|}{|K|}}\chi _{\Delta }(g_{i}gg_{i}^{-1})
\label{charinduce}
\end{equation}%
where $\chi _{\Delta }$ is the character of the representation $\Delta $ .

A very important fact is that in the case of semi-direct product groups $G=%
\mathit{A}\circledS \mathit{B}$ with the $\mathit{A}$ component being an
abelian invariant subgroup, this method when used with Wigner's and Mackey
little group and orbit techniques, provides with the complete set of
inequivalent unitary irreps \ of the group,(c.f. \cite{corn}) faithful or
non faithful. In using these techniques one determines all the non
isomorphic semi-direct product subgroups $K$ of $G$, whose unitary
inequivalent irreps will be used to induce the corresponding irreps of $G$
according to the above theorem.

In particular, consider the vector space of characters of $\mathit{A}$ with
basis the irreducible characters $\chi _{\mathit{A}}^{j}$. The action of $%
\mathit{B}$ on the space of characters given by $g_{B}\chi _{\mathit{A}%
}^{j}(a)=\chi _{\mathit{A}}^{j}(g_{B}ag_{B}^{-1})=$ $\chi _{\mathit{A}%
}^{j^{\prime }}(a^{\prime })$ where $g_{B}\in \mathit{B}$, $a,a^{\prime }\in 
\mathit{A}$, provide in general a reducible representation of $\mathit{B}$.
To each $\chi _{\mathit{A}}^{j}$ the set of elements of $\mathit{B}$ that
leave $\chi _{\mathit{A}}^{j}$ invariant form a subgroup of $\mathit{B}$
called the little group $\mathit{B}_{j}$ and the action of the coset $%
\mathit{B/}$ $\mathit{B}_{j}$ on $\chi _{\mathit{A}}^{j}$ defined the orbit $%
\mathit{Orb}(\chi _{\mathit{A}}^{j})$. Varying $\chi _{\mathit{A}}^{j}$ be
obtain all the orbits which in general are not distinct. In what will
follows below we shall determine all the distinct orbits of $\mathit{B}$. In
order to apply the above theorem 1 we choose the subgroups of $G$ which have
the form $K_{j}=\mathit{A}\circledS \mathit{B}_{j}$ where $\mathit{B}_{i}$
are subgroup of $\mathit{B}$, whose elements leave each character $\chi _{%
\mathit{A}}^{j}$ , representative of the distinct orbits of $\mathit{A}$
invariant. Finally we construct the irreps of each $K_{j}$ which will be
used to induce the irreps of $G$.

Turning to our case where $G=HW_{2^{s}}=\mathit{H}\circledS \mathit{B}$
where $\mathit{H}$ and $\mathit{B}$ are as in( \ref{defH}) (\ref{defB})
respectively.

The space of the unitary inequivalent irreps of the invariant abelian
subgroup $\mathit{H}$ $\approx 
\mathbb{Z}
_{2^{s}}\times 
\mathbb{Z}
_{2^{s}}$\ is given by the set $\ \Omega =\{\chi _{\mathit{H}%
}^{(p,q)},\;p,q=0,1,...,2^{s}-1\}$ of characters, where 
\begin{equation}
\chi _{\mathit{H}}^{(p,q)}(z^{m}x^{n})=e^{\frac{2\pi i}{2^{s}}(mp+nq)},\;
\label{irrepsH}
\end{equation}%
for all $m,\;n,\;p,\;q=0,1,...,2^{s}-1$.

The action of $\mathit{B}$ on the characters results as follows ,

\begin{equation}
y\chi _{\mathit{H}}^{(p,\;q)}(z^{m}x^{n})=\chi _{\mathit{H}%
}^{(p,q)}(yz^{m}x^{n}y^{-1})=\chi _{\mathit{H}}^{(p,\;p+q)}(z^{m}x^{n})
\label{actionb}
\end{equation}

for every $y\in \mathit{B.}$This action can be visualized as the geometrical
action of the group $\mathit{B}$ on the two dimensional discrete torus $%
\mathit{T}_{2}(2^{s})=$ $%
\mathbb{Z}
_{2^{s}}\times 
\mathbb{Z}
_{2^{s}}$. Every point on this torus ($m,n$) is mapped to point ($m+1,n$).
That is the action of B is an elementary translation across the first
component. In the dual torus ($p,q$) which is the space of the characters$%
\chi _{\mathit{H}}^{(p,\;q)}$, this action induces the mapping ($p,q$)$%
\rightarrow $($p,p+q$), (\ref{actionb}).

The little groups $\mathit{B}^{(p,q)}$ are defined as the subgroups of $%
\mathit{B}$ that leave invariant the point of the dual torus ($p,q$). This
implies that $\mathit{B}^{(p,q)}={y^{l}\in \mathit{B}: lp=0(mod2^{s})}$. In
order to determine the elements of $\mathit{B}^{(p,q)}$ we distinguish two
cases: for the case $p\neq 0$ and $p=0$. In the case $p\neq 0$ we extract
the maximum power of 2, $p=2^{t}u_{t}(mod2^{s})$, introducing the parameters 
$t$ and $u_{t}$. In this case we obtain 
\begin{equation}
\mathit{B}^{(p,q)}=\{(y^{2^{s-t}})^{v},\;v=0,1,...,2^{t}-1\}\approx 
\mathbb{Z}
_{2^{t}}  \label{littlep}
\end{equation}%
where $u_{t}=1,3,5..,2^{s-t}-1,\;$and $t=0,1,...,s-1$. In the second case $%
p=0,$ we easily obtain $\mathit{B}^{(0,q)}=\mathit{B}$.

We observe that for fixed $t$, $\mathit{B}^{(p,q)}$ is independent of the
values of $u_{t}$ and $q$. Thus there are $2^{2s-t-1}$ identical little
groups as in (\ref{littlep}). Similarly for $p=0$, there are $2^{s}$
identical ones.

The inequivalent irreps, $\Gamma _{\mathit{B}^{(p,q)}}^{r}$ , of $\mathit{B}%
^{(p,q)}$ labeled by $r$ , which will be needed in what follows, are all
one-dimensional. As $\mathit{B}^{(p,q)}$ are cyclic, these are given by 
\begin{equation}
\Gamma _{\mathit{B}^{(p,q)}}^{r}(\left( y^{2^{s-t}}\right) ^{v}) = e^{\frac{%
2\pi i}{2^{t}}rv},\;r,v=0,1,...,2^{t}-1,\;p\neq 0  \label{littlereps}
\end{equation}
\begin{equation}
\Gamma _{\mathit{B}^{(0,q)}}^{r}(y^{l}) = e^{\frac{2\pi i}{2^{s}}rl},
\;r,l=0,1,...,2^{s}-1, \;p=0  \label{littlereps2}
\end{equation}

Turning now to the structure of the orbit of $\chi _{\mathit{H}}^{(p,\;q)}$, 
$\mathit{Orb}(\chi _{\mathit{H}}^{(p,q)})$, under $\mathit{B}$, notice that
since the action of the elements of $\mathit{B}^{(p,q)}$ leave $\chi _{%
\mathit{H}}^{(p,\;q)}$ invariant, $\mathit{Orb}(\chi _{\mathit{H}}^{(p,q)})$
consists of the irreducible characters of $\mathit{B}$ obtained by the
action of the factor group $\mathit{B}/\mathit{B}^{(p,q)}=\{y^{k},%
\;k=0,1,...,2^{s-t}-1\}$ on $\chi _{\mathit{H}}^{(p,\;q)}$ . Using (\ref%
{actionb}) and (\ref{irrepsH}) $\mathit{Orb}(\chi _{\mathit{H}}^{(p,q)})$
are given by

\begin{equation}
\mathit{Orb}(\chi _{\mathit{H}}^{(p,q)}) = \{\chi _{\mathit{H}%
}^{(p,\;kp+q)},\;k=0,1...,2^{s-t}-1\},\;p\neq 0
\end{equation}
\begin{equation}
\mathit{Orb}(\chi _{\mathit{H}}^{(0,q)}) = \{\chi _{\mathit{H}%
}^{(0,\;q)}\},\;p=0  \label{orbpq}
\end{equation}
In appentix A we show that all the distinct orbits we are interested in, are
produced by \textit{constraining} the label $q$ in the range $%
q=0,1...,2^{t}-1,$ when $p=2^{t}u_{t}(mod2^{s})$, while when $p=0$, $%
\;q=0,1...,2^{s}-1.$ Counting arguments show that there are $2^{s}$ orbits
of the form $\mathit{Orb}(\chi _{\mathit{H}}^{(0,q)})$ and for each value of 
$t$, $2^{s-t-1}\times 2^{t}$ orbits of the form $\mathit{Orb}(\chi _{\mathit{%
H}}^{(2^{t}u_{t},q)})$, the total number of distinct orbits being $%
\sum_{t=0}^{s-1}(2^{s-t-1}2^{t})+2^{s}=\allowbreak 2^{s-1}s+2^{s}$ and as
expected their union contains the\ $%
\sum_{t=0}^{s-1}(2^{s-t-1}2^{t}2^{s-t})+2^{s}=\allowbreak 2^{2s}$ elements
of the character set of $\mathit{H}$. For each distinct orbit we choose its
representative to be $\chi _{\mathit{H}}^{(p,q)}$ for $p\neq 0$ or $\chi _{%
\mathit{H}}^{(0,\;q)}$.

Now for each ($p,q$) labeling the representative of a distinct orbit, we
construct the invariant subgroup of $HW_{2^{s}}$, $\mathit{K}^{(p,q)}=%
\mathit{H}\circledS \mathit{B}^{(p,q)}$ which is given by 
\begin{equation}
\mathit{K}^{(p,q)}=\{z^{m}x^{n}y^{2^{s-t}v},\;m,n=0,...,2^{s}-1,%
\;v=0,...,2^{t}-1\}  \label{Kgroup}
\end{equation}%
has order $|\mathit{K}^{(p,q)}|=2^{2s+t}$and the coset space $HW_{2^{s}}/%
\mathit{K}^{(p,q)}=\mathit{B}/\mathit{B}^{(p,q)}$.

The next step in the construction is the determination of the one
dimensional unitary representations of $\mathit{K}^{(p,q)}$, $\Gamma _{%
\mathit{K}^{(p,q)}}^{(p,q),r}$, obtained by tensoring the unitary irreps of $%
\mathit{B}^{(p,q)}$ and $\mathit{H}$. For each unitary irrep of $\mathit{B}%
^{(p,q)}$, $\Gamma _{\mathit{B}^{(p,q)}}^{r}$, given by (\ref{littlereps})
and each distinct orbit representative $\chi _{\mathit{H}}^{(p,q)}$ , we
find that 
\begin{equation}
\Gamma _{\mathit{K}^{(p,q)}}^{(p,q),r}(z^{m}x^{n}y^{2^{s-t}v}) = \chi _{%
\mathit{H}}^{(p,q)}(z^{m}x^{n})\Gamma _{\mathit{B}^{(p,q)}}^{r}(y^{2^{s-t}v})%
\text{, }p\neq 0  \label{irrepT}
\end{equation}
\begin{equation}
\Gamma _{\mathit{K}^{(0,q)}}^{(0,q),r}(z^{m}x^{n}y^{l}) = \chi _{\mathit{H}%
}^{(0,q)}(z^{m}x^{n})\Gamma _{\mathit{B}^{(0,q)}}^{r}(y^{l})\;p=0
\end{equation}
where $m,n,l=0,1,...,2^{s}-1$ . More explicitly, 
\begin{equation}
\Gamma _{\mathit{K}^{(p,q)}}^{(p,q),r}(z^{m}x^{n}y^{2^{s-t}v}) = e^{\frac{%
2\pi i}{2^{s}}(mp+nq)}e^{\frac{2\pi i}{2^{t}}rv}, \;v=0,...,2^{t}-1
\label{irrepT1}
\end{equation}
\begin{equation}
\Gamma _{\mathit{K}^{(0,q)}}^{(0,q),r}(z^{m}x^{n}y^{l}) = e^{\frac{2\pi i}{%
2^{s}}(nq+rl)}  \label{irrepT2}
\end{equation}

We are ready now to apply \textit{theorem 1}, by making the identifications $%
G\equiv HW_{2^{s}}$, $K\equiv \mathit{K}^{(p,q)}$ and $\Delta \equiv \Gamma
_{\mathit{K}^{(p,q)}}^{(p,q),r}$ . In this way we get the complete set of
unitary irreps of $HW_{2^{s}}$ which are presented by the following theorem
(c.f. \cite{corn}) :

\textit{Theorem 2}. The matrix representations $\Gamma ^{(p,q),r}$ of $%
HW_{2^{s}}$ induced by the those of $\mathit{K}^{(p,q)}$, i.e. $\Gamma
^{(p,q),r}=\Gamma _{\mathit{K}^{(p,q)}}^{(p,q),r}\uparrow HW_{2^{s}}$,
provide with the complete set of inequivalent unitary irreducible matrix
representations of $HW_{2^{s}}$ and they are given by 
\begin{equation}
\Gamma ^{(p,q),r}(z^{m}x^{n}y^{l})_{kj}=\left\{ 
\begin{array}{c}
\Gamma _{\mathit{K}^{(p,q)}}^{(p,q),r}(y^{k}z^{m}x^{n}y^{l}y^{-j})\;\ \text{%
for all\ \ }y^{k}y^{l}y^{-j}\in \mathit{B}^{(p,q)} \\ 
0\text{\ \ \ \ \ \ otherwise}%
\end{array}%
\right\} \text{\ }  \label{ind1}
\end{equation}%
%
%
%
%
%
%
%
%
%
%
%
%
where $k,\;j=0,1...,2^{s-t}-1$ and for all $m,n,l=0,1,...,2^{s}-1$.
Explicitely, upon using (\ref{irrepsH}) and (\ref{irrepT2}), (\ref{ind1}) we
find that 
\begin{eqnarray}
\Gamma ^{(p,q),r}(z^{m}x^{n}y^{l})_{kj} &=&\sum_{v=0}^{2^{t}-1}\chi _{%
\mathit{H}}^{(p,\;kp+q)}(z^{m}x^{n})\chi _{\mathit{B}%
^{(p,q)}}^{r}(y^{2^{s-t}v})\delta _{k+l-j,2^{s-t}v}  \notag \\
&=&\sum_{v=0}^{2^{t}-1}e^{\frac{2\pi i}{2^{s}}pm}e^{\frac{2\pi i}{2^{s}}%
(kp+q)n}e^{\frac{2\pi i}{2^{t}}(rv)}\delta _{k+l-j,2^{s-t}v}  \label{ind2}
\end{eqnarray}%
%
%
%
%
%
%
%
%
%
%
%
%
and the corresponding characters are given by 
\begin{eqnarray}
\chi ^{(p,q),r}(z^{m}x^{n}y^{l}) &=&  \notag \\
&&\sum_{k=0}^{2^{s-t}-1}\sum_{v=0}^{2^{t}-1}\chi _{\mathit{H}%
}^{(p,pk+q)}(z^{m}x^{n})\chi _{\mathit{B}^{(p,q)}}^{r}(y^{l})\delta
_{l,2^{s-t}v}\   \label{charind2}
\end{eqnarray}%
for all $m,n,l=0,1,...,2^{s}-1$ and where $\delta _{k+l-j,2^{s-t}v}=1$ iff $%
k+l-j=2^{s-t}v(mod2^{s})$.

\textit{Proof }

Unitarity of $\Gamma ^{(p,q),r}$ stems from the unitarity of $\Gamma _{%
\mathit{K}^{(p,q)}}^{(p,q),r}$. Irreducibility is easily proved by ckecking
that $\sum_{m,n,l=0}^{2^{s}-1}\left\vert \chi
^{(p,q),r}(z^{m}x^{n}y^{l})\right\vert ^{2}$ equals the order of $HW_{2^{s}}$%
. Indeed, using (\ref{charind2}) we have that 
\begin{eqnarray*}
\sum_{m,n,l=0}^{2^{s}-1}\left\vert \chi
^{(p,q),r}(z^{m}x^{n}y^{l})\right\vert ^{2} \\
=\sum_{m,n,l=0}^{2^{s}-1}\chi ^{(p,q),r}(z^{m}x^{n}y^{l})\chi
^{(p,q),r}(z^{m}x^{n}y^{l})^{\ast } \\
= \sum_{m,n,l=0}^{2^{s}-1}(\sum_{k,k^{\prime
}=0}^{2^{s-t}-1}\sum_{v,v^{\prime }=0}^{2^{t}-1}\chi _{\mathit{H}%
}^{(p,pk+q)}(z^{m}x^{n})\chi _{\mathit{B}^{(p,q)}}^{r}(y^{l})\delta
_{l,2^{s-t}v}\chi _{\mathit{H}}^{(p,pk^{\prime }+q)}(z^{m}x^{n})^{\ast }\chi
_{\mathit{B}^{(p,q)}}^{r}(y^{l})^{\ast }\delta _{l,2^{s-t}v^{\prime }}) \\
= \sum_{m,n,l=0}^{2^{s}-1}(\sum_{k,k^{\prime
}=0}^{2^{s-t}-1}\sum_{v,v^{\prime }=0}^{2^{t}-1}e^{\frac{2\pi i}{2^{s}}mp}e^{%
\frac{2\pi i}{2^{s}}(pk+q)n}e^{\frac{2\pi i}{2^{t}}rv}\delta
_{l,2^{s-t}v}e^{-\frac{2\pi i}{2^{s}}mp}e^{-\frac{2\pi i}{2^{s}}(pk^{\prime
}+q)n}e^{-\frac{2\pi i}{2^{t}}rv^{\prime }}\delta _{l,2^{s-t}v^{\prime }}) \\
=2^{2s}2^{s-t}\sum_{l=0}^{2^{s}-1}\sum_{v,v^{\prime }=0}^{2^{t}-1}e^{\frac{%
2\pi i}{2^{t}}r(v-v^{\prime })}\delta _{l,2^{s-t}v}\delta
_{l,2^{s-t}v^{\prime }} =2^{2s}2^{s-t}\sum_{v,v^{\prime }=0}^{2^{t}-1}e^{%
\frac{2\pi i}{2^{t}}r(v-v^{\prime })}\delta _{v,v^{\prime }}=2^{3s}
\end{eqnarray*}%
%
%
%
%
%
%
%
%
%
%
%
%
%
%
%
%
To show the completeness of the set of the unitary irreps thus constructed
we have first to demonstrate if we chose any other character $\chi _{\mathit{%
H}}$ than the representative $\chi _{\mathit{H}}^{(p,q)}$ from a distinct
orbit we obtain equivalent representations and secondly that the same is
true if we chose any other character from other non distinct orbits as
classified above in appendix 1. These will be investigated by using the well
known fact that for two representations $\Gamma ^{(p,q),r}$ and $\Gamma
^{(p^{\prime },q^{\prime }),r\prime }$ to be equivalent a necessary and
sufficient condition is the equality of their character systems. The
demonstration of the completeness is finalized by showing that the sum of
the square of the dimensions of all the constructed irreps equals $2^{3s}$ ,
the order of group.

To this end, let $\chi ^{(p,q),r}=\chi _{\mathit{H}}^{(p,q)}\chi _{\mathit{B}%
^{(p,q)}}^{r}$ be the character of an irrep constructed above where $\chi _{%
\mathit{H}}^{(p,q)}$ is the character representative of the corresponding
distinct orbit and let $\chi ^{(p,kp+q),r^{\prime }}=\chi _{\mathit{H}%
}^{(p,kp+q)}\chi _{\mathit{B}^{(p,kp+q)}}^{r^{\prime }}$be a\ character
where $\chi _{\mathit{H}}^{(p,kp+q)}$ is in the orbit of $\chi ^{(p,q)}$ for
some fixed $k=1,2,...,2^{s-t}-1.$ Since by construction $\mathit{B}^{(p,q)}=%
\mathit{B}^{(p,kp+q)}$, taking $r=r^{\prime }$ and using (\ref{charind2}),
with $p=2^{t}u_{t}$, $\chi ^{(p,q),r}$ is given by: 
\begin{eqnarray}
\chi ^{(p,q),r}(z^{m}x^{n}y^{l}) =\sum_{k^{\prime
}=0}^{2^{s-t}-1}\sum_{v=0}^{2^{t}-1}e^{\frac{2\pi i}{2^{s}}pm}e^{\frac{2\pi i%
}{2^{s}}(k^{\prime }p+q)n}e^{\frac{2\pi i}{2^{t}}(rv)}\delta _{l,2^{s-t}v} 
\notag \\
=2^{s-t}\sum_{v=0}^{2^{t}-1}e^{\frac{2\pi i}{2^{s}}pm}e^{\frac{2\pi i}{2^{s}}%
qn}\delta _{u_{t}n,0}e^{\frac{2\pi i}{2^{t}}(rv)}\delta _{l,2^{s-t}v}
\label{app1}
\end{eqnarray}%
%
%
%
%
%
%
%
%
%
%
%
%
%
%
%
%
%
%
%
%
%
%
%
%
%
while for $\chi ^{(p,kp+q),r^{\prime }}$ 
\begin{eqnarray}  \label{app2}
\chi ^{(p,kp+q),r^{\prime }}(z^{m}x^{n}y^{l}) &=&\sum_{k^{\prime \prime
}=0}^{2^{s-t}-1}\sum_{v=0}^{2^{t}-1}e^{\frac{2\pi i}{2^{s}}pm}e^{\frac{2\pi i%
}{2^{s}}[(k^{\prime \prime }+k)pn+qn]}e^{\frac{2\pi i}{2^{t}}(rv)}\delta
_{l,2^{s-t}v}  \notag \\
&=&2^{s-t}\sum_{v=0}^{2^{t}-1}e^{\frac{2\pi i}{2^{s}}pm}e^{\frac{2\pi i}{%
2^{s}}(kp+q)n}\delta _{u_{t}n,0}e^{\frac{2\pi i}{2^{t}}(rv)}\delta
_{l,2^{s-t}v}  \notag \\
&=&2^{s-t}\sum_{v=0}^{2^{t}-1}e^{\frac{2\pi i}{2^{s}}pm}e^{\frac{2\pi i}{%
2^{s-t}}ku_{t}n}e^{\frac{2\pi i}{2^{s}}qn}\delta _{u_{t}n,0}e^{\frac{2\pi i}{%
2^{t}}(rv)}\delta _{l,2^{s-t}v}  \notag \\
\end{eqnarray}
for all $m,n,l=0,1,...,2^{s}-1$. Since $\delta _{u_{t}n,0}$ is evaluated mod$%
2^{s-t}$, for $m=l=0$, the non zero characters are those for which $%
n=0,2^{s-t},2\times 2^{s-t},3\times 2^{s-t}....$. Thus comparison of (\ref%
{app1}) with (\ref{app2}) shows that $\chi ^{(p,q),r}(z^{m}x^{n}y^{l})=\chi
^{(p,kp+q),r}(z^{m}x^{n}y^{l})$ for all $k=0,1,...,2^{s-t}-1$ and so $\Gamma
^{(p,q),r}$ and $\Gamma ^{(p,kp+q),r}$ are equivalent.

Turning now to the case of two representations $\Gamma ^{(p,q),r}$ and $%
\Gamma ^{(p,q+j),r^{\prime }}$ where $q$ takes a fixed value between $0$ to $%
2^{t}-1$ and $j$ is as in \ref{distinctK} . The character $\chi
^{(p,q+j),r^{\prime }}=\chi _{\mathit{H}}^{(p,q+j)}\chi _{\mathit{B}%
^{(p,q+j)}}^{r^{\prime }}$ \ is such that $\chi _{\mathit{H}}^{(p,q+j)}$ is
taken to be an orbit representative of the non distinct orbit $\mathit{Orb}%
(\chi _{\mathit{H}}^{(p,q+j)})$. Again since by construction $\mathit{B}%
^{(p,q)}=\mathit{B}^{(p,q+j)}$, taking $r=r^{\prime }$ and using (\ref%
{charind2}), with $p=2^{t}u_{t}$, $\chi ^{(p,q),r}$ is given by (\ref{app1})
while $\chi ^{(p,q+j),r}$ is given by: 
\begin{eqnarray}
\chi ^{(p,q+j),r}(z^{m}x^{n}y^{l})
&=&\sum_{k=0}^{2^{s-t}-1}\sum_{v=0}^{2^{t}-1}e^{\frac{2\pi i}{2^{s}}pm}e^{%
\frac{2\pi i}{2^{s}}(kp+q+j)n}e^{\frac{2\pi i}{2^{t}}(rv)}\delta
_{l,2^{s-t}v}  \notag \\
&=&2^{s-t}\sum_{v=0}^{2^{t}-1}e^{\frac{2\pi i}{2^{s}}pm}e^{\frac{2\pi i}{%
2^{s}}(q+j)n}\delta _{u_{t}n,0}e^{\frac{2\pi i}{2^{t}}(rv)}\delta
_{l,2^{s-t}v}  \label{app3}
\end{eqnarray}
where again for $m=l=0$, the non zero characters are those for which $n$ is
a multiple of $2^{s-t}$. Since $j$ is a multiple of $2^{t}$ the term $e^{%
\frac{2\pi i}{2^{s}}jn}=1$ which implies the equality of(\ref{app1}) and (%
\ref{app3}) and thus the equivalence of $\Gamma ^{(p,q),r}$ and $\Gamma
^{(p,q+j),r}$.

Finally for the unitary irreps thus constructed, observe that for each value
of $t=0,1,...,s-1$, there correspond $2^{s-t-1}$ $\times 2^{2t}$ irreps $%
\Gamma ^{(2^{t}u_{t},q),r}$ , all of the same dimensionality $2^{s-t}$ and
for $p=0$, there correspond $2^{2s}$ $1$-dimensional irreps $\Gamma
^{(0,q),r}$. Thus the sum of the square of the dimensions of all these
irreps is $\sum_{t=0}^{s-1}\left( 2^{s-t-1}2^{2t}\right)
2^{2s-2t}+2^{2s}=\allowbreak 2^{3s}$ as is the order of $HW_{2^{s}}$. 
$\Diamond $

For $t=0$, the matrix irreps given by $\Gamma ^{(u_{0},0),0}$, $%
u_{0}=1,3,...,2^{s}-1$, are faithful $2^{s}$-dimensional, for $t=1,...s-1$,
the $\Gamma ^{(2^{t}u_{t},q),r}$, $u_{t}=1,3,...,2^{s-t}-1$, $%
q,r=0,1,...,2^{t}-1$, irreps are non-faithful $2^{s-t}$ -dimensional and for 
$p=0$, the $\Gamma ^{(0,q),r}$ , $q,r=0,1,...,2^{s}-1$ irreps are\ $1$-dim.
Counting arguments show that the number N$_{s}$ of inequivalent irreps of $%
HW_{2^{s}}$ is given by

\begin{eqnarray*}
N_{s} &=&\sum_{t=0}^{s-1}\left( 2^{s-t-1}2^{2t}\right)
+2^{2s}=2^{s-1}(3\times 2^{s}-1) \\
&=&\sum_{t=0}^{s-1}N_{s}(t)+2^{2s}
\end{eqnarray*}%
where $N_{s}(t)$ is the number of inequivalent irreps of dimension $%
d=2^{s-t} $ and the number of inequivalent $1$-dimensional irreps is equal
to $2^{2s}$. As expected $N_{s}$ gives precisely the same result with the
number of conjugacy classes $N_{C}$ in (\ref{numcon}). In the Table I the
numbers $N_{s}$ of irreps (and conjugacy classes) of $HW_{2^{s}}$\ for
various values of $s$ are shown: 
\begin{table}[tbp]
\caption{ Numbers $N_{s}$ of irreps (and conjugacy classes) of $HW_{2^{s}}$
for various values of $s$}%
\begin{tabular}{|l|l|l|l|l|l|l|l|l|l|l|}
\hline
$s$ & $1$ & $2$ & $3$ & $4$ & $5$ & $6$ & $7$ & $8$ & $9$ & $10$ \\ \hline
$N_{s}$ & $5$ & $22$ & $92$ & $376$ & $1520$ & $6112$ & $24512$ & $98176$ & $%
392960$ & $1572352$ \\ \hline
\end{tabular}%
\end{table}

In closing this section we present a simplification of notation to be used
in the rest of the article. A general element of the group will be denoted
by $J_{mnl}$ $\equiv z^{m}x^{n}y^{l}$, a unitary matrix irrep. $\Gamma
^{(p,q),r}$ , by $D\equiv \lbrack (p,q),r]$ so that $J_{mnl}^{D}\equiv
\Gamma ^{(p,q),r}(J_{mnl})$. Thus, for the \textit{non} trivial irreps (i.e. 
$p\neq 0$) the commutator of two elements $J_{mnl}^{D},$ $J_{m^{\prime
}n^{\prime }l^{\prime }}^{D}$ is given by: 
\begin{equation}
\lbrack J_{mnl}^{D},J_{m^{\prime }n^{\prime }l^{\prime }}^{D}]=(\omega
_{s-t}^{u_{t}n^{\prime }l}-\omega _{s-t}^{u_{t}nl^{\prime }})J_{m+m^{\prime
},n+n^{\prime },l+l^{\prime }}^{D}  \label{comj}
\end{equation}%
where $\omega _{s-t}=e^{\frac{2\pi i}{2^{s-t}}}$ $\ m,n,l=0,1,...,2^{s}-1$.

\section{Generalized Finite Fourier Transforms}

In this section we investigate the existence of a transformation between $x$
and $y$ and the conditions under which this transformation can be a Finite
Fourier Transform. In Finite quantum mechanics $x$ represents the position
operator in the diagonal form and $y$ the shift operator of a fictitious
particle moving on a discrete circle with $2^{s-t}$ equidistant points. The
corresponding phase space is the discrete torus $T_{2}[2^{s-t}]=%
\mathbb{Z}
_{2^{s-t}}\times 
\mathbb{Z}
_{2^{s-t}}$ .

To obtain the matrix form of such a Finite Fourier Transform we shall first
present the matrix form of the generators $z$, $x$ and $y$ of $HW_{2^{s}}$.
These matrix forms are important for explicit calculations in problems of
finite quantum mechanics.

In the representations constructed above , it is understood that the carrier
space of the irreps $\Gamma ^{(p,q),r}$ can be taken to be 
\begin{equation}
V^{D}\equiv V^{(p,q),r}=span\{|j>,\;j=0,1,...,2^{s-t}-1\}  \label{car1}
\end{equation}%
where we use the canonical basis in the Dirac notation, $|j>=(e_{j})_{k}=%
\delta _{kj}$ and $<k|j>=\delta _{kj}$ . Following (\ref{ind2}), the action
of $z,\;x,\;y$ elements on $V^{D}$, is given by 
\begin{equation}
z_{D}|j>=\omega _{s-t}^{u_{t}}\delta _{k,j}|k>,\;x_{D}|j>=\omega
_{s}^{q}\omega _{s-t}^{u_{t}k}\delta _{k,j}|k>  \label{carr}
\end{equation}%
\begin{eqnarray}
y_{D}|0 &>&=\omega _{t}^{r}|2^{s-t}-1>,  \notag \\
y_{D}|j &>&=\delta _{k+1,j}|k>,\;j=1,...,2^{s-t}-1.  \label{oqpD}
\end{eqnarray}

Thus the $2^{s-t}\times 2^{s-t}$ matrix form of $z_{D}$,$\;x_{D}$, and $%
y_{D} $ are given by 
\begin{equation}
z_{D} = \omega _{s-t}^{u_{t}}I_{2^{s-t}},\;x_{D}=\omega _{s}^{q}\left[ 
\begin{array}{cccc}
1 & 0 & 0 & 0 \\ 
0 & \omega _{s-t}^{u_{t}} & \cdots & 0 \\ 
\vdots & \ddots & \ddots & \vdots \\ 
0 & \ldots & 0 & \omega _{s-t}^{u_{t}\left( 2^{s-t}-1\right) }%
\end{array}%
\right] ,\;  \label{matrixrep}
\end{equation}
\begin{equation}
y_{D} = \left[ 
\begin{array}{cccccc}
0 & 1 & 0 & ... & ... & 0 \\ 
0 & 0 & 1 & 0 & ... & 0 \\ 
\vdots & \ddots & \ddots & \ddots & \ddots & \vdots \\ 
0 & 0 & \ddots & 0 & \ddots & 0 \\ 
0 & 0 & 0 & \ddots & 0 & 1 \\ 
\omega _{t}^{r} & 0 & 0 & 0 & 0 & 0%
\end{array}%
\right]
\end{equation}

Moreover using (\ref{matrixrep}) it is shown that 
\begin{equation}
y_{D}x_{D}=z_{D}x_{D}y_{D}  \label{basic}
\end{equation}%
which can be used to show that 
\begin{equation}
x_{D}^{2^{s-t}}=\omega _{t}^{q}I_{2^{s-t}},\;y_{D}^{2^{s-t}}=\omega
_{t}^{r}I_{2^{s-t}},\;x_{D}^{2^{s}}=y_{D}^{2^{s}}=z_{D}^{2^{s}}=I_{2^{s}}
\label{period}
\end{equation}

Relations (\ref{period})\ show that besides the expected $2^{s}$
-periodicity, the non faithful irreps ($t\neq 0$) satisfy twisted boundary
conditions for the motions around the $x$ and $y$ cycles of phase space
torus $T_{2}[2^{s-t}]$, characterized by powers of $\omega _{t}$. From the
explicit matrix form of $x$ and $y$ (\ref{matrixrep}), we observe that the
fictitious particle is desrcibed by quantum mechanical states carrying
winding number $u_{t}$ and satisfying the twisted boundary conditions with
phase $\omega _{t}^{q}$ for the rotation around the $x$-axis and \ $\omega
_{t}^{r}$ for the $y$ axis. It follows that the corresponding Fourier
Transform is expected to be modified with respect to the standard one.

Using (\ref{matrixrep}), the normalized eigenvectors and corresponding
eigenvalues $\lambda $ of $y_{D}$, solutions of the equation $\lambda
^{2^{s-t}}=\omega _{t}^{r}$ , are respectively given by 
\begin{eqnarray}
|\psi _{k} &>&=\frac{1}{\sqrt{2^{s-t}}}\left[ 
\begin{array}{ccccc}
1 & \omega _{s}^{r}\omega _{s-t}^{k} & \omega _{s}^{2r}\omega _{s-t}^{2k} & 
\cdots & \omega _{s}^{(2^{s-t}-1)r}\omega _{s-t}^{(2^{s-t}-1)k}%
\end{array}%
\right] ^{T}  \notag \\
\lambda _{k} &=&\omega _{s}^{r}\omega _{s-t}^{k},\;k=0,...,2^{s-t}-1,
\label{eigy}
\end{eqnarray}%
where $|\psi _{k}>$ denotes the kth column eigenvector in the Dirac
notation. 
Then the diagonalizing matrix $F_{D}$ of $y_{D}$, can be easily deduced to
have the product form 
\begin{equation}
F_{D}=\Omega _{r}F_{s-t}  \label{ffd}
\end{equation}%
where the matrix elements of $\Omega _{r}$, $F_{s-t}$ and $F_{D}$ are
respectively given by 
\begin{eqnarray}
(\Omega _{r})_{kj} &=&\omega _{s}^{rk}\delta _{kj},  \notag \\
(F_{s-t})_{kj} &=&\frac{1}{\sqrt{2^{s-t}}}\omega _{s-t}^{kj} \\
(F_{D})_{kj} &=&\omega _{s}^{rk}\omega _{s-t}^{kj}  \label{mffd}
\end{eqnarray}%
with $k,j=0,...,2^{s-t}-1$. In particular the matrices $\Omega _{r}$ and $%
F_{s-t}$ have the matrix form 
\begin{eqnarray}
\Omega _{r} &=&diag\left[ 1,\omega _{s}^{r},\omega _{s}^{2r},...\omega
_{s}^{(2^{s-t}-1)r}\right] ,\; \\
F_{s-t} &=&\frac{1}{\sqrt{2^{s-t}}}\left[ 
\begin{array}{ccccc}
1 & 1 & 1 & 1 & 1 \\ 
1 & \omega _{s-t} & \omega _{s-t}^{2} & \cdots & \omega _{s-t}^{(2^{s-t}-1)}
\\ 
1 & \omega _{s-t}^{2} & \omega _{s-t}^{4} & \cdots & \omega
_{s-t}^{2(2^{s-t}-1)} \\ 
\vdots & \vdots & \vdots & \cdots & \vdots \\ 
1 & \omega _{s-t}^{(2^{s-t}-1)} & \omega _{s-t}^{2(2^{s-t}-1)} & \cdots & 
\omega _{s-t}^{(2^{s-t}-1)(2^{s-t}-1)}%
\end{array}%
\right]
\end{eqnarray}%
%
%
%
%
%
%
%
%
%
%
%
%
and $F_{s-t}$ is the standard Finite Fourier Transform for $s-t$ qubits. In
the discrete phase space $F_{s-t}$ \ represents the rotation by $\pi /2$
degrees i.e. $F_{s-t}^{4}=I_{D}$. Moreover $\left(
F_{D}y_{D}F_{D}^{-1}\right) ^{u_{t}}=\omega _{s}^{ru_{t}}\omega
_{s}^{-q}x_{D}^{-1}$ and thus 
\begin{equation}
F_{D}y_{D}^{u_{t}}F_{D}^{-1}=\omega _{s}^{ru_{t}-q}x_{D}^{-1}
\end{equation}%
where $u_{t}=1,3,...2^{s-t}-1$ and $r,q=0,...,2^{t}-1$.

\section{The irreducible characters and fussion rules of $HW_{2^{s}}$ irreps.%
}

Up on using (\ref{irrepsH}), (\ref{littlereps}) , (\ref{ind2}) and (\ref%
{matrixrep}), the characters $\chi ^{D}(J_{mnl})=trJ_{mnl}^{D}$ of the $%
2^{s-t}$ -dim. representations are given by 
\begin{eqnarray}
\chi ^{D}(J_{mnl}) &=&\sum_{k=0}^{2^{s-t}-1}\chi _{\mathit{H}%
}^{(p,pk+q)}(J_{mn0})\chi _{\mathit{B}^{(p,q)}}^{r}(J_{00l})\delta
_{l,2^{s-t}v} \\
&=&\omega _{s-t}^{u_{t}m}\text{\ }\sum_{k=0}^{2^{s-t}-1}\sum_{v=0}^{2^{t}-1}%
\omega _{s-t}^{u_{t}kn}\omega _{s}^{qn}\omega _{t}^{rv}\delta _{l,2^{s-t}v}
\label{charind}
\end{eqnarray}%
for all $m,n,l=0,1,...,2^{s}-1$. \ In particular, the \textit{non-zero}
values of characters are given by 
\begin{equation}
\chi ^{D}(J_{m,2^{s-t}v_{1},2^{s-t}v_{2}})=2^{s-t}\omega
_{s-t}^{u_{t}m}\omega _{t}^{v_{1}q+v_{2}r}  \label{nonzerochar}
\end{equation}%
where $m=0,1,...,2^{s}-1$, and $v_{1},v_{2}=0,..,2^{t}-1$. For example in
the case of characters of $HW_{4}$ i.e. $s=2$, for we have 22 inequivalent
irreps: (i) for $t=0$ , the two $4$-dim faithful irreps $\Gamma
^{(1,0),0},\;\Gamma ^{(3,0),0}$, (ii) for $t=1$ , the four $2$-dim
non-faithful, $\Gamma ^{(2,q),r},\;q,r=0,1$ and (iii) for $t=2$, the sixteen 
$1$-dim, $\Gamma ^{(0,q),r},\;q,r=0,1,2,3$.

In the following we shall establish the fussion algebra of the unitary
irreps of $HW_{2^{s}}$ as constructed above. Given any two such irreps, $%
D_{i}$, $D_{j}$, the fussion algera is defined by 
\begin{equation}
D_{i}\otimes D_{j}=\sum_{k}N_{ij}^{k}D_{k}
\end{equation}%
where $N_{ij}^{k}$ is the multiplicity of the $D_{k}$ irrep appearing in the
decomposition of the above tensor product. Since these multiplicities
completely specifies the fussion algebra we shall provide with the explicit
relation which computes them. For simplicity of exposition, let $%
D_{1},\;D_{2}$ be \textit{any two} given unitary irreps of $HW_{2^{s}}$ and
let $D_{3}$ be some irrep that would appear in the decomposition of $%
D_{1}\otimes D_{2}$ . Implementing the well known relation of the
multiplicities in the case of the tensor product of two finite group irreps
(c.f. \cite{corn}) we obtain 
\begin{equation}
N_{12}^{3}=\frac{1}{2^{3s}}\sum_{m,n,l=0}^{2^{s}-1}\chi
^{D_{1}}(J_{m,n,l})\chi ^{D_{2}}(J_{m,n,l})\chi ^{D_{3}}(J_{m,n,l})^{\ast }
\label{number}
\end{equation}%
where the $\chi ^{D_{i}}$, $i=1,2,3$ are given by (\ref{charind}). Using (%
\ref{charind})and (\ref{nonzerochar}), relation (\ref{number}) becomes 
\begin{equation}
N_{12}^{3}=2^{s-t_{2}+t_{1}-t_{3}}\delta
_{p_{1}+p_{2},\;p_{3}(mod2^{s})}\delta
_{q_{1}+q_{2},\;q_{3}(mod2^{t_{1}})}\delta
_{r_{1}+r_{2},\;r_{3}(mod2^{t_{1}})}  \label{numberfuse}
\end{equation}%
%
%
%
%
%
%
%
%
%
%
%
%
where $p_{1}=2^{t_{1}}u_{t_{1}}$, $p_{2}=2^{t_{2}}u_{t_{2}}$, $%
p_{3}=2^{t_{3}}u_{t_{3}}$ and without loss of generality it has been assumed
that $t_{1}\leq t_{2}$. Consequently the fussion algebra of $HW_{2^{s}}$ is
completely specified by 
\begin{eqnarray}
\lbrack (p_{1},q_{1}),r_{1}]\otimes \lbrack (p_{2},q_{2}),r_{2}] &\approx & 
\notag \\
\oplus
_{q_{3}(mod2^{t_{1}})=q_{1}+q_{2}(mod2^{t_{1}}),r_{3}(mod2^{t_{1}})=r_{1}+r_{2}(mod2^{t_{1}})}2^{s-t_{2}+t_{1}-t_{3}}[(p_{1}+p_{2},q_{3}),r_{3}]
\label{fuse}
\end{eqnarray}%
%
%
%
%
%
%
%
%
%
%
%
%
It should be understood that in the above multiplicity and fussion formula
the values of $q_{3}$ and $r_{3}$ are in the range $\{0,1,...,2^{t_{3}}-1\}$
such that $q_{3}(mod2^{t_{1}})=(q_{1}+q_{2})(mod2^{t_{1}})$ and $%
r_{3}(mod2^{t_{1}})=(r_{1}+r_{2})(mod2^{t_{1}})$ and where $t_{3}$ is
obtained by $p_{3}=(p_{1}+p_{2})(mod2^{s})=2^{t_{3}}u_{t_{3}}$. In
particular, in the case of the fussion of two identical (up to equivalence)
irreps $[(p,q),r]\otimes \lbrack (p,q),r]$ where $p=2^{t}u_{t}$ we obtain
that 
\begin{eqnarray}
\lbrack (p,q),r]\otimes \lbrack (p,q),r] &\approx &  \notag \\
\oplus
_{q_{3}(mod2^{t})=2q(mod2^{t}),%
\;r_{3}(mod2^{t})=2r(mod2^{t})}2^{s-t-1}[(2^{t+1}u_{t+1},q_{3}),r_{3}]
\label{fuse2}
\end{eqnarray}%
%
%
%
%
%
%
%
%
%
%
%
%
where with $t\neq 0$, we have set $%
p_{3}=(2p)(mod2^{s})=(2^{t+1}u_{t})(mod2^{s})=2^{t+1}u_{t+1}$ and $%
q_{3},r_{3}\in \{0,...,2^{t+1}-1\}$ such that $q_{3}=2q(mod2^{t})$, $%
r_{3}=2r(mod2^{t})$.

In the case of the fussion of two faithfull irreps $[(p_{1},0),0]\otimes
\lbrack (p_{2},0),0]$ where $p_{1},p_{2}$ are odd we obtain that 
\begin{equation}
\lbrack (p_{1},0),0]\otimes \lbrack (p_{2},0),0]\approx \oplus _{
q_{3}(mod1)=0 \newline
r_{3}(mod1)=0}2^{s-t_{3}}[(p_{3},q_{3}),r_{3}]  \label{fuse3}
\end{equation}
with $p_{3}=(p_{1}+p_{2})(mod2^{s})=2^{t_{3}}u_{t_{3}}$ and all $%
q_{3},r_{3}\in \{0,1,...,2^{t_{3}}-1\}$ from which its is apparent that only
non-faithful irreps participate in the decomposition.

\subsection{Example 1. Fussion rules of $HW_{2}$}

According to the analysis of the previous sections, the inequivalent irreps
of $HW_{2}$ are: (i) for $t=0$ , the one $2$-dim faithful irreps $\Gamma
^{(1,0),0}$, (ii) for $t=1$ , the four $1$-dim non-faithful, $\Gamma
^{(0,q),r},\;q,r=0,1$. Using (\ref{fuse}), the set of fussion rules among
the irreps are given by 
\begin{eqnarray*}
\Gamma ^{(1,0),0}\otimes \Gamma ^{(1,0),0} &\approx &\Gamma ^{(0,0),0}\oplus
\Gamma ^{(0,0),1}\oplus \Gamma ^{(0,1),0}\oplus \Gamma ^{(0,1),1} \\
\Gamma ^{(1,0),0}\otimes \Gamma ^{(0,q),r} &\approx &\Gamma ^{(1,0),0}\text{
\ \ \ for all \ }q,r=0,1 \\
\Gamma ^{(0,0),0}\otimes \Gamma ^{(0,q),r} &\approx &\Gamma ^{(0,q),r}\text{
\ \ \ for all \ }q,r=0,1 \\
\Gamma ^{(0,0),1}\otimes \Gamma ^{(0,0),1} &\approx &\Gamma
^{(0,0),0},\;\Gamma ^{(0,0),1}\otimes \Gamma ^{(0,1),0}\approx \Gamma
^{(0,1),1} \\
\Gamma ^{(0,0),1}\otimes \Gamma ^{(0,1),1} &\approx &\Gamma
^{(0,1),0},\;\Gamma ^{(0,1),0}\otimes \Gamma ^{(0,1),0}\approx \Gamma
^{(0,0),0} \\
\Gamma ^{(0,1),0}\otimes \Gamma ^{(0,1),1} &\approx &\Gamma
^{(0,0),1},\;\Gamma ^{(0,1),1}\otimes \Gamma ^{(0,1),1}\approx \Gamma
^{(0,0),0}
\end{eqnarray*}

\subsection{Example 2. Fussion rules of $HW_{2^{2}}$}

According to the analysis of the previous sections, the inequivalent irreps
of $HW_{2^{2}}$ are: (i) for $t=0$ , the two $4$-dim faithful irreps $\Gamma
^{(1,0),0},\;\Gamma ^{(3,0),0}$, (ii) for $t=1$ , the four $2$-dim
non-faithful, $\Gamma ^{(2,q),r},\;q,r=0,1$ and (iii) for $p=0$ (or
equivalently for $t=2$), the sixteen $1$-dim, $\Gamma
^{(0,q),r},\;q,r=0,1,2,3$. Using (\ref{fuse}), the set of fussion rules
among the non-trivial irreps are given by 
\begin{eqnarray*}
\Gamma ^{(1,0),0}\otimes \Gamma ^{(1,0),0} &\approx &\Gamma
^{(3,0),0}\otimes \Gamma ^{(3,0),0} \\
&\approx &\oplus _{q,r=0,1}2\Gamma ^{(2,q),r} \\
\Gamma ^{(1,0),0}\otimes \Gamma ^{(3,0),0} &\approx &\oplus
_{q,r=0..3}\Gamma ^{(0,q),r} \\
\Gamma ^{(1,0),0}\otimes \Gamma ^{(2,q),r} &\approx &2\Gamma ^{(3,0),0}\text{
\ \ \ for all \ }q,r=0,1 \\
\Gamma ^{(3,0),0}\otimes \Gamma ^{(2,q),r} &\approx &2\Gamma ^{(1,q),r}\text{
\ for all \ }q,r=0,1 \\
\Gamma ^{(2,0),0}\otimes \Gamma ^{(2,0),0} &\approx &\Gamma
^{(2,0),1}\otimes \Gamma ^{(2,0),1}\approx \Gamma ^{(2,1),0}\otimes \Gamma
^{(2,1),0} \\
&\approx &\Gamma ^{(2,1),1}\otimes \Gamma ^{(2,1),1}\approx \oplus
_{q,r=0,2}\Gamma ^{(0,q),r} \\
\Gamma ^{(2,0),0}\otimes \Gamma ^{(2,0),1} &\approx &\Gamma
^{(2,1),0}\otimes \Gamma ^{(2,1),1} \\
&\approx &\Gamma ^{(0,0),1}\oplus \Gamma ^{(0,0),3}\oplus \Gamma
^{(0,2),1}\oplus \Gamma ^{(0,2),3} \\
\Gamma ^{(2,0),0}\otimes \Gamma ^{(2,1),0} &\approx &\Gamma ^{(0,1),0}\oplus
\Gamma ^{(0,1),2}\oplus \Gamma ^{(0,3),0}\oplus \Gamma ^{(0,3),2} \\
\Gamma ^{(2,0),0}\otimes \Gamma ^{(2,1),1} &\approx &\Gamma
^{(2,0),1}\otimes \Gamma ^{(2,1),0}\approx \oplus _{q,r=1,3}\Gamma ^{(0,q),r}
\\
\Gamma ^{(2,0),1}\otimes \Gamma ^{(2,1),1} &\approx &\Gamma ^{(0,1),0}\oplus
\Gamma ^{(0,1),2}\oplus \Gamma ^{(0,3),0}\oplus \Gamma ^{(0,3),2} \\
\Gamma ^{(2,1),0}\otimes \Gamma ^{(2,1),1} &\approx &\Gamma ^{(0,0),1}\oplus
\Gamma ^{(0,0),3}\oplus \Gamma ^{(0,2),1}\oplus \Gamma ^{(0,2),3}
\end{eqnarray*}

\section{Conclusion}

We conclude this work by a summary of our results. We have introduced a
definition of the Heisenberg Weyl group $HW_{2^{s}}$ using only a
presentation of the relations for its generators. This presentation allows a
generalization of the standard discrete $HW_{2^{s}}$ group where only the
primitive root of unity $\omega =e^{\frac{2\pi i}{2^{s}}}$ appears as the
generator of its center. Our generalization allows all the possible roots of
unity to appear with the crucial result the appearance of a rich spectrum of
non faithful representations. We have thus to apply the Wigner-Mackey method
of orbits and little groups to this extended group. Indeed we have found
that all the representations are classified by a triplet of integers$(p,q,r)$%
, \ taking appropriate values in the range from $0$ to $2^{s}-1$. The
dimensions of these representations are $2^{s-t}$ where $t$ runs from $0$ to 
$s$. When $t=0$ the representations are faithful while for $t$ from $1$ to $%
s $ the representations are non-faithful. For each of these representations
we have determined explicitly their matrix form as well as their characters
and their fussion rules. In particular the rich structure of the fussion
rules provide decompositions with non trivial multiplicities which may be of
interest in anyonic systems. We have examined also the relation between the
"position" and "momentum" operators for any irrep with the result the
existence of generalized finite Fourier transforms.

To discuss possible applications and future work we note that for
applications in quantum circuits it is necessary to construct the
corresponding quantum circuit for the unitary matrices of any representation
and any group element. To this end, it is enough to construct the quantum
circuits for the "momentum" and "position" matrices $P$ and $Q$.

For applications in the area of quantum Hall effect, we should consider an
important class of group elements of $HW_{2^{s}}$, the so called magnetic
translations which represent the motion of an electron in two dimensional
toroidal lattice $%
\mathbb{Z}
_{2^{s}}\times 
\mathbb{Z}
_{2^{s}}$ in the presence of a transverse magnetic flux $2\pi /2^{s}$ per
lattice plaquete, in units of quantum of magnetic flux \cite{zak}, \cite%
{flo3} and references therein. The problem which appears for this value of
flux is that the magnetic translations are ill defined and one has to go to
a new phase space lattice $%
\mathbb{Z}
_{2^{2s}}\times 
\mathbb{Z}
_{2^{2s}}$ and the extension of the $HW_{2^{s}}$ to the tensor product $%
HW_{2^{s}}\otimes HW_{2^{s}}$. The solution to this problem will appear in
later article by the authors. 

In this area considerations on the construction of models of topological
quantum computations lead to the study of new excitations of many electron
systems the so called anyonic excitations of exotic fractional statistics 
\cite{kitaev}. Here the representations of the finite Heisenberg group $%
HW_{2^{s}}$ as well as its automorphism group play an important role in the
study of so called metaplectic anyons \cite{nayak2}. The Heisenberg group
for a system of N electrons moving on the above toroidal lattice is the
N-fold tensor product of $HW_{2^{s}}$ and its automorphism group is the
finite symplectic group $Sp_{2N}(%
\mathbb{Z}
_{2^{s}})$ so we have to construct quantum circuits for the elements of this
group. The anyonic excitations are eigenstates of the corresponding unitary
operators because of the relation of the braid group $B_{2N+1}$ and the
symplectic group. So in future work we shall study the metaplectic
representation of $Sp_{2N}(%
\mathbb{Z}
_{2^{s}})$.

\appendix

\section{ Determination of distinct orbits}

Consider an irreducible character $\chi _{\mathit{H}}^{(p,\;q)}$ of $\mathit{%
B}$ and the corresponding little group $\mathit{B}^{(p,q)}$, where the
values of $p$ are given either by $p=2^{t}u_{t}$, for each $t=0,1...,s-1$
and each $u_{t}=1,3,...2^{s-t}-1$ or by $p=0$ and in both cases the values
of $q=0,1...,2^{s}-1$. Since the action of the elements of $\mathit{B}%
^{(p,q)}$ leave $\chi _{\mathit{H}}^{(p,\;q)}$ invariant, the orbit of $\chi
_{\mathit{H}}^{(p,\;q)}$, $\mathit{Orb}(\chi _{\mathit{H}}^{(p,q)})$, under $%
\mathit{B}$ consists of the irreducible characters of $\mathit{B}$ obtained
by the action of the factor group $\mathit{B}/\mathit{B}^{(p,q)}=\{y^{k},%
\;k=0,1,...,2^{s-t}-1\}$ on $\chi _{\mathit{H}}^{(p,\;q)}$ given by

\begin{equation}
y^{k}\chi _{\mathit{H}}^{(p,\;q)}(z^{m}x^{n})=\chi _{\mathit{H}%
}^{(p,q)}(y^{k}z^{m}x^{n}y^{-k})=\chi _{\mathit{H}}^{(p,\;kp+q)}(z^{m}x^{n})
\end{equation}
where (\ref{irrepsH}) and (\ref{actionb}) have been used. That is

\begin{eqnarray}
\mathit{Orb}(\chi _{\mathit{H}}^{(p,q)}) &=&\{\chi _{\mathit{H}%
}^{(p,\;kp+q)},\;k=0,1...,2^{s-t}-1\},\;p\neq 0  \notag \\
\mathit{Orb}(\chi _{\mathit{H}}^{(0,q)}) &=&\{\chi _{\mathit{H}%
}^{(0,\;q)}\},\;p=0  \label{orbit}
\end{eqnarray}%
It is obvious that the sets $\mathit{Orb}(\chi _{\mathit{H}}^{(0,q)})$ for $%
q\in \{0,1,...,2^{s}-1\}$, have all one element and are all distinct.
However this is not the case for $\mathit{Orb}(\chi _{\mathit{H}}^{(p,q)})$%
,\ $p\neq 0$ since for fixed $p$, $kp+q$ is evaluated $mod2^{s}$ . For fixed 
$p=2^{t}u_{t}$ the sets $K_{q}=\{kp+q,\;k=0,1...,2^{s-t}-1\}$ take the form $%
K_{q}=\{q,2^{t}u_{t}+q,2\times 2^{t}u_{t}+q,...,(2^{s-t}-1)2^{t}u_{t}+q\}$
from which it is obvious that changing the values of $u_{t}$ $\in
\{1,3,...,2^{s-t}-1\}$ leads to a mere rearrangement of the elements of the
set. Thus it suffices to find the distinct sets $K_{q}$ when $u_{t}=1$,
given by

\begin{equation}
K_{q}=\{q,2^{t}+q,2\times 2^{t}+q,...,(2^{s-t}-2)2^{t}+q,(2^{s-t}-1)2^{t}+q\}
\label{set1}
\end{equation}%
when $q$ takes the values from $0$ to $2^{s}-1$. For $q\in
\{0,1,...,2^{t}-1\}$\ it is evident that the sets $K_{q}$ are all distinct
since for the maximum value $q=2^{t}-1$, the maximum value in the set is $%
2^{s}-1$. For $q=2^{t}$ , $2^{t}+1,...,2\times 2^{t}-1$ direct evaluation
mod2$^{s}$ of the elements in the (\ref{set1}) shows that $K_{2^{t}}=K_{0}$, 
$K_{2^{t}+1}=K_{1}$, $K_{2^{t}+2}=K_{2}$,..., $K_{2\times
2^{t}-1}=K_{2^{t}-1}$. Similarly for $q=2\times 2^{t}$ , $2\times
2^{t}+1,...,3\times 2^{t}-1$ it is shown that $K_{2\times 2^{t}}=K_{0}$, $%
K_{2\times 2^{t}+1}=K_{1}$, ..., $K_{3\times 2^{t}-1}=K_{2^{t}-1}$.
Continuing this evaluation up to the values of $q=(2^{s-t}-1)\times 2^{t}$ , 
$(2^{s-t}-1)\times 2^{t}+1$, $(2^{s-t}-1)\times 2^{t}+2$, ... ,$%
(2^{s-t}-1)\times 2^{t}+$ $2^{t}-1$ we finally obtain that for each fixed
pair $s,t$, $K_{0}$, $K_{1}$, ..., $K_{2^{t}-1}$ are all distinct while for
the rest values of $q$ from $2^{t}$ to $2^{s}-1$, it is deduced that

\begin{eqnarray}
K_{j+q} &=&K_{q},\;\text{for each \ }q=0,1,...,2^{t}-1  \notag \\
\text{and all \ }j &=&2^{t},2\times 2^{t},3\times 2^{t},...,(2^{s-t}-1)2^{t}
\label{distinctK}
\end{eqnarray}

Consequently the distinct orbits are those for which either $p=0$ and $%
q=0,1...,2^{s}-1$ or $p=2^{t}u_{t}$ and $q=0,1,...,2^{t}-1$. These
participate in the construction of the complete set of inequivalent irreps
of $HW_{2^{s}}$. Counting arguments show that the number of distinct orbits $%
N_{o}=\sum_{t=0}^{s-1}\left( 2^{s-t-1}2^{t}\right) +2^{s}=2^{s-1}(s+2).$

Closing this appendix we can illustrate the determination of distinct orbits
with the example of $HW_{2^{2}}$. In this case $s=2$ and $t=0,1$ and $%
q=0,1,2,3$. For $p=0$ we have 4 distinct orbits $\mathit{Orb}(\chi _{\mathit{%
H}}^{(0,0)})$,$\mathit{Orb}(\chi _{\mathit{H}}^{(0,1)})$,$\mathit{Orb}(\chi_{%
\mathit{H}}^{(0,2)})$, $\mathit{Orb}(\chi_{\mathit{H}}^{(0,3)})$ of one
element each. For $t=0$, $u_{0}=1,3$, $p=u_{0}=1,3$ and for q=0,1,2,3 we
obtain

\begin{eqnarray}
\mathit{Orb}(\chi _{\mathit{H}}^{(1,0)}) =\{\chi _{\mathit{H}}^{(1,0)},\chi
_{\mathit{H}}^{(1,1)},\chi _{\mathit{H}}^{(1,2)},\chi _{\mathit{H}}^{(1,3)}\}
\notag \\
=\mathit{Orb}(\chi _{\mathit{H}}^{(1,1)})=\mathit{Orb}(\chi _{\mathit{H}%
}^{(1,2)})=\mathit{Orb}(\chi _{\mathit{H}}^{(1,3)})
\end{eqnarray}
\begin{eqnarray}
\mathit{Orb}(\chi _{\mathit{H}}^{(3,0)}) =\{\chi _{\mathit{H}}^{(3,0)},\chi
_{\mathit{H}}^{(3,3)},\chi _{\mathit{H}}^{(3,2)},\chi _{\mathit{H}}^{(3,1)}\}
\notag \\
=\mathit{Orb}(\chi _{\mathit{H}}^{(3,1)})=\mathit{Orb}(\chi _{\mathit{H}%
}^{(3,2)})=\mathit{Orb}(\chi _{\mathit{H}}^{(3,3)})
\end{eqnarray}
so the distinct orbits can be taken to be $\mathit{Orb}(\chi _{\mathit{H}%
}^{(1,0)})$ and $\mathit{Orb}(\chi _{\mathit{H}}^{(3,0)})$, that is for $q=0$%
. For t=1, $u_{1}$=1, so $p=2u_{1}=2,$ and for q=0,1,2,3 we obtain

\begin{eqnarray*}
\mathit{Orb}(\chi _{\mathit{H}}^{(2,0)}) &=&\{\chi _{\mathit{H}%
}^{(2,0)},\chi _{\mathit{H}}^{(2,2)}\}=\mathit{Orb}(\chi _{\mathit{H}%
}^{(2,2)}) \\
\mathit{Orb}(\chi _{\mathit{H}}^{(2,1)}) &=&\{\chi _{\mathit{H}%
}^{(2,1)},\chi _{\mathit{H}}^{(2,3)}\}=\mathit{Orb}(\chi _{\mathit{H}%
}^{(2,3)})
\end{eqnarray*}%
so the distinct orbits can be taken to be $\mathit{Orb}(\chi _{\mathit{H}%
}^{(2,0)})$ and $\mathit{Orb}(\chi _{\mathit{H}}^{(2,1)})$, that is for $%
q=0,1$.


\begin{thebibliography}{99}
\bibitem{woot} Gibbons, K. S., Hoffman, M. J., Wooters, W. K.:Discrete phase
space based on finite fields. Phys. Rev. A \textbf{70}, 62101 (2004).

\bibitem{woot2} Wooters, W. K.: Picturing qubits in phase space. IBM J. Res.
Dev. \textbf{48}, 99 (2004).

\bibitem{shor} Calderbank, A. R., Shor, P. W.: Good quantum error-correcting
codes exist. Phys. Rev. A \textbf{54}, 1098 (1996).

\bibitem{howe} Howe, R.:Nice error bases, mutually unbiased bases, induced
representations, the Heisenberg group and finite geometries. Indag. Math.
(N.S.) \textbf{16}, 553 (2005).

\bibitem{Banks} Banks, T., Fischler, W.: The holographic spacetime model of
cosmology. Int.J.Mod.Phys.D \textbf{27}, 1846005 (2018).

\bibitem{caroll} Bao, N., Carroll, S. M., Singh, A.: The Hilbert Space of
Quantum Gravity Is Locally Finite-Dimensional. Int. J. Mod. Phys. D, \textbf{%
26} 1743013 (2017).

\bibitem{carroll2} Singh, A., Carroll, S. M.: Modeling Position and Momentum
in Finite-Dimensional Hilbert Spaces via Generalized Clifford Algebra.
quant-ph 1806.10134v1.(2018)

\bibitem{weyl} Weyl, H.: The Theory of Groups and Quantum Mechanics. Dover
Books on Mathematics, Dover Publications (1950).

\bibitem{swinger} Schwinger, J.: Proceedings of the National Academy of
Sciences \textbf{46} 257; 550; 883;1401 (1960).

\bibitem{balian} Balian, R., Itzykson, C.: Observations on finite quantum
mechanics. C. R. Acad. Sc. Paris \textbf{303}, 773 (1986).

\bibitem{vourdas} Vourdas, A.:Quantum Systems with Finite Hilbert Space.
Rep. Prog. Phys., \textbf{67}, 267 (2004).

\bibitem{jagan} Jaganathan, R.: The Legacy of Alladi Ranakrishnan in the
Mathematical Sciences. Eds. K. Alladi, J. R. Klauder and C. R. Rao Springer
(2010) DOI 10.1007/978-1-4419-6263-8\_22.

\bibitem{flo8} Athanasiu, G. G., Floratos, E. G.: Coherent states in Finite
Quantum Mechanics. Nuclear Physics B, \textbf{425}, 343 (1994).

\bibitem{zak} Zak, J. : Weyl-Heisenberg group and magnetic translations in
finite phase space. Phys. Rev. B, \textbf{39}, 694 (1989).

\bibitem{grass} Grassberger, J., Hormann, G.: A note on representations of
the finite Heisenberg group and sums of greatest common divisors. Discrete
Mathematics and Theoretical Computer Sience, \textbf{4}, 91 (2001).

\bibitem{roditi} Altschuler, D., Beran, P., Lacki,J., Roditi, I.: Twisted
vertex operators and representations of finite Heisenberg groups. Internat.
J. Modern Phys. A \textbf{4}, 921 (1989).

\bibitem{misag} Misaghian, M.: The representations of the Heisenberg group
over a finite field. Armen. J. Math. \textbf{3}, 162 (2010).

\bibitem{schulte} Schulte, J. : Harmonic analysis on finite Heisenberg
group. European J. Combin., \textbf{25}, 3273 (2004).

\bibitem{gure} Gurevich, S., Hadani, R.: The Weil representation in
characteristic two. Advances in Mathematics, \textbf{230}, 894 (2012).

\bibitem{apleby} Appleby,R. M. : Properties of the extended Clifford group
with applications to SIC-POVMS and MUBS. quant-ph 09095233.


\bibitem{mack} Mackey, G.W.: Induce representation of Groups and Quantum
Mechanics. W. A. Benjamin, NY (1968).

\bibitem{corn} Cornwell, J. F.: Group Theory in Physics, Vol1 Academic
Press, Chap. 5, pp. 118-127, 2nd ed. (1984)

\bibitem{flo3} Floratos, E. G., Nicolis, S.: An SU(2) analogue of the
Azbel-Hofstadter Hamiltonia. J. Phys. A Math. Gen., \textbf{31}, 3961 (1998).

\bibitem{kitaev} Kitaev, A. Y.: Fault-tolerant quantum computation by
anyons. Annals of Physics, \textbf{303}, 2 (2003).

\bibitem{nayak2} Hastings, M. B., Nayak, C, Wang, Z.: Metaplectic anyons,
Majorana zero modes, and their computational power. Physical Rev. B, \textbf{%
87}, 165421 (2013).
\end{thebibliography}
\end{document}